\documentclass[journal,10pt]{IEEEtran}

\usepackage{cite}
\usepackage{epsfig}
\usepackage{epstopdf}
\usepackage{graphicx}
\usepackage[dvipsnames]{xcolor}
\usepackage{tikz}
\usetikzlibrary{spy,backgrounds,arrows,positioning,shapes.geometric,circuits.logic.US}
\usepackage{pgfplots}
\usepackage{multirow}
\usepackage{upgreek}
\usepackage{amssymb}
\usepackage{amsmath}
\usepackage{cases}
\usepackage{url}
\usepackage{pifont}
\usepackage{ctable}
\usepackage{bm}
\usepackage{amsthm}

\usepackage{tabularx}
\usepackage{booktabs}
\usepackage{filecontents}
\usepackage{subcaption}
\newcolumntype{Y}{>{\centering\arraybackslash}X}

\theoremstyle{definition} 
\theoremstyle{definition} 
\theoremstyle{definition} 
\theoremstyle{definition}

\usepackage[linesnumbered, ruled, vlined, noend]{algorithm2e}

\usepackage{algorithmic}

\usepackage{array}

\newcommand{\fixme}[2]{\ifx&#2&{\leavevmode\color{red}#1}\else{\leavevmode\color{red}FIXME\{}#1{\leavevmode\color{red}\}}\footnote{{\leavevmode\color{red}#2}}\PackageWarning{Fixme}{#1: #2}\fi}

\newcommand{\newstuff}[2]{\ifx&#2&{\leavevmode\color{blue}#1}\else{\leavevmode\color{blue}FIXME\{}#1{\leavevmode\color{blue}\}}\footnote{{\leavevmode\color{blue}#2}}\PackageWarning{Newstuff}{#1: #2}\fi}

\DeclareMathOperator*{\lg2}{log}

\title{A fixed latency ORBGRAND decoder architecture with LUT-aided error-pattern scheduling}

\author{\IEEEauthorblockN{Carlo~Condo,~\IEEEmembership{Senior~Member,~IEEE}\\}
\IEEEauthorblockA{Infinera Corporation\\
Email: ccondo@infinera.com}} 

\begin{document}

\maketitle
\begin{abstract}

Guessing Random Additive Noise Decoding (GRAND) is a universal decoding algorithm that has been recently proposed as a practical way to perform maximum likelihood decoding. 
It generates a sequence of possible error patterns and applies them to the received vector, checking if the result is a valid codeword.
Ordered reliability bits GRAND (ORBGRAND) improves on GRAND by considering soft information received from the channel. 
Both GRAND and ORBGRAND have been implemented in hardware, focusing on average performance, sacrificing worst case throughput and latency. 
In this work, an improved pattern schedule for ORBGRAND is proposed. 
It provides $>0.5$ dB gain over the standard schedule at a block error rate $\le 10^{-5}$, and outperforms more complex GRAND flavors with a fraction of the complexity.
The proposed schedule is used within a novel code-agnositic decoder architecture: the decoder guarantees fixed high throughput and low latency, making it attractive for latency-constrained applications. 
It outperforms the worst-case performance of decoders by orders of magnitude, and outperforms many best-case  figures.
Decoding a code of length 128, it achieves a throughput of $79.21$ Gb/s with $58.49$ ns latency, yielding better energy efficiency and comparable area efficiency with respect to the state of the art.
\end{abstract}


\IEEEpeerreviewmaketitle

\section{Introduction} \label{sec:intro}

Maximum Likelihood (ML) decoding is an optimal decoding approach that can be applied to virtually any code. 
It foresees the comparison of the vector to be decoded with all the possible codewords in the codebook, and the selection of the codeword that minimizes the error probability as the decoded vector.
The complexity of ML decoding is very high, making it impractical for many applications.

Guessing Random Additive Noise Decoding (GRAND) \cite{GRAND_first} is a recently-proposed algorithm that can perform ML or near-ML decoding with limited complexity, and that can be used to decode any type of code. 
It has been shown to work well with short, high-rate codes. 
Unlike decoding algorithms that use the structure of the code to detect and correct errors in the received vector, GRAND attempts to guess the error pattern that was applied on the transmitted codeword.
ML decoding is achieved by scheduling the attempted error patterns in descending order of likelihood: here lies the inherent challenge of GRAND decoding, that has led to various incarnations being proposed \cite{SGRAND_first,soft_GRAND,ORBGRAND_first}. 
They evolve on the original premise by making use of soft information received from the channel to better infer the error-pattern schedule. 
Ordered Reliability Bits GRAND (ORBGRAND) \cite{ORBGRAND_first} sorts the channel soft information in order of reliability, and schedules the error patterns based on their logistic weights.
Thanks to its good trade-off between performance and complexity, it has attracted the interest of the research community, and improvements to ORBGRAND have been recently proposed \cite{iLWO,LGRAND}.

The attractiveness of GRAND-based algorithms has led to the proposal of decoder hardware implementations \cite{GRAND_arch,GRANDchip,ORBARCH_J}.
The standard GRAND is used in \cite{GRAND_arch,GRANDchip}, while \cite{ORBARCH_J} implements ORBGRAND.
To limit the implementation complexity, the scheduled error patterns are constrained in terms of Hamming weight $HW$ and, in case of \cite{ORBARCH_J}, logistic weight $LW$ as well.
These decoders focus on average performance; they potentially require a large number of patterns to achieve good error-correction performance, but the decoding process is on average very short, since it is usually sufficient to attempt the decoding with few, highly-likely error patterns to obtain a valid codeword.
These design choices lead to high average throughput and short average latency, but cause the worst case scenarios to have very bad performance.
This is a noticeable hurdle towards the widespread acceptance of GRAND-based decoding, given the strict latency and demanding performance constraints of recent and upcoming communication standards, like the ultra-reliable low-latency scenario foreseen in the 5G standard \cite{3GPP_R15,URLLC,URLLC2}. 

In this work, the aforementioned problem is tackled on two sides.
As a first step, the core issue of GRAND-based decoding of scheduling the error patterns in the most effective order is addressed. 
A high-performance error-pattern schedule is proposed, that enables ORBGRAND to provide its best error-correction performance in literature with a small maximum number of error patterns.
This schedule further improves the performance of the one proposed in \cite{iLWO}, through the addition of a number of empirically-observed highly likely patterns as the first ones attempted; it yields more than 0.5 dB gain over the original schedule \cite{ORBGRAND_first} at a block error rate $\le 10^{-5}$.
The proposed schedule can be easily used by any ORBGRAND decoder architecture in literature.
Secondly, a code-agnostic decoder architecture is proposed, that can decode any binary linear code with code length $\le N$ and code rate $\ge R_{min}$.
Enabled by the characteristics of the newly proposed schedule, it adopts a different paradigm than the architectures in literature, and guarantees a high fixed throughput and very low fixed latency. 
The variable number of patterns required to achieve successful decoding is leveraged to reduce power consumption instead of increasing average throughput.
The decoder has been synthesized in 7 nm FinFET technology with various sets of design parameters, and shown to yield up to $79.21$ Gb/s throughput and downto $40.58$ ns latency.

The remainder of the paper is organized as follows. 
Section \ref{sec:prel} introduces GRAND-based decoding and the relevant evolutions.
The proposed error-pattern schedule is detailed in Section \ref{sec:LAGRAND}, and simulation results are presented.
The decoder architecture is described in Section \ref{sec:arch}, whereas implementation results and comparison with the state of the art are presented in Section \ref{sec:impl}.
Finally, Section \ref{sec:conc} draws the conclusion.

\section{Preliminaries} \label{sec:prel}




%
Let us define a binary linear block code code $\mathcal{C}$, identified by the $k \times n$ generator matrix $\mathbf{G}$ and the $(n-k) \times n$ parity check matrix $\mathbf{H}$.
Then, a source vector $\mathbf{u}$ of $k$ bits is encoded into a codeword $\mathbf{x}$ of $n$ bits through $\mathbf{x}=\mathbf{u}\cdot \mathbf{G}$.
The $2^k$ possible $\mathbf{x}$ compose the codebook of $\mathcal{C}$, for which the following is true:
\begin{equation}
\forall \mathbf{x} \in \mathcal{C},\mathbf{H} \cdot \mathbf{x}^{\rm T} = \mathbf{0}~,
\end{equation}
where $\mathbf{0}$ is the all-zero vector. 
Let $\mathbf{x}$ be transmitted over a noisy channel; given the received soft-value vector $\mathbf{y}$, the hard-decided vector inferred from $\mathbf{y}$ can be expressed as ${\rm HD}(\mathbf{y}) = \mathbf{x} \oplus \mathbf{e}$, where $\mathbf{e}$ is the error pattern applied by the channel.
In case $\mathbf{H} \cdot {\rm HD}(\mathbf{y})^{\rm T} \neq \mathbf{0}$, errors have been detected.
In the remainder of this work, the elements of $\mathbf{y}$ are supposed to be logarithmic likelihood ratios (LLRs).

The Guessing Random Additive Noise Decoding (GRAND) \cite{GRAND_first} algorithm attempts to find the error pattern $\mathbf{e}$ applied by the channel.
After generating a test pattern $\mathbf{e}$, it computes ${\rm HD}(\mathbf{y}) \oplus \mathbf{e}$.
The codebook is queried by computing $\mathbf{H} \cdot ({\rm HD}(\mathbf{y}) \oplus \mathbf{e})^{\rm T}$: in case the result is $\mathbf{0}$, the decoding is successful, and vector $\hat{\mathbf{y}}={\rm HD}(\mathbf{y}) \oplus \mathbf{e}$ is returned, otherwise a new $\mathbf{e}$ is generated.
While these steps should be repeated until a valid codeword is found, GRAND with abandonment (GRANDAB, \cite{GRAND_first}) foresees the termination of the decoding process after a maximum amount of codebook queries $Q_{max}$ has been performed. 

To achieve ML decoding, the error patterns should be scheduled from the most probable to the least probable one. 
The scheduling of the error patterns is at the core of GRAND-based decoding design, and has a large impact on the algorithm error-correction performance. 
The optimal ordering for binary symmetric channels is based on the Hamming weight $HW$ of $\mathbf{e}$, but this schedule is strongly suboptimal for additive white Gaussian noise channels (AWGN). 
The Ordered Reliability Bits GRAND (ORBGRAND) \cite{ORBGRAND_first} algorithm is a better suited option for complex channels. 
By making use of the soft information vector $\mathbf{y}$ instead of ${\rm HD}(\mathbf{y})$ only, it infers a refined error-pattern schedule.
The elements of vector $\mathbf{y}$ are sorted in ascending order of reliability, i.e. in ascending order of magnitude in case of LLRs, resulting in the index permutation $\pi$ and the sorted vector $\pi(\mathbf{y})$. 
Error patterns $\mathbf{e}$ are applied to $\pi(\mathbf{y})$ in ascending \emph{logistic weight order} (LWO). 
Given the ordered vector $\mathbf{v} = (v_0,\dots,v_{HW-1})$ containing the indices of the nonzero entries of $\mathbf{e}$, and its length $HW$, then the logistic weight $LW$ of $\mathbf{e}$ can be computed as
\begin{equation}\label{eq:LW}
LW(\mathbf{e}) = \sum_{i=0}^{HW-1} (v_i+1)~.
\end{equation}
Error patterns with the same $LW$ can be scheduled in any order. 
The syndrome calculation can be then performed in natural or permuted order:
\begin{align}
&\mathbf{H}\cdot({\rm HD}(\mathbf{y})\oplus\pi^{-1}(\mathbf{e}))^T~,\nonumber \\
&\pi(\mathbf{H})\cdot(\pi({\rm HD}(\mathbf{y}))\oplus \mathbf{e})^T~, \nonumber
\end{align}
where $\pi(\mathbf{H})$ is the column-permuted $\mathbf{H}$.
In Section \ref{sec:arch}, the permuted order syndrome $\mathbf{s}=\pi(\mathbf{H})\cdot\mathbf{z}^T$ is used, where $\mathbf{z}=\pi({\rm HD}(\mathbf{y}))\oplus \mathbf{e}$.

GRAND-based decoding is universal, as it can be applied to any code.
In the remainder of the paper, two particular code types are used as examples: Bose-Chaudhuri-Hocquenghem (BCH) codes \cite{BCH_forney} and polar codes \cite{polar}.
BCH codes are widespread block codes used in a variety of practical applications, including communications, storage and cryptography. 
They feature a very flexible code design, that can cater to different error correction needs through the tuning of the code minimum distance.
Polar codes \cite{polar} are binary block codes based on the polarization property of their kernel matrix. 
They have been the object of intense research in the last decade, and they have been included in the 5G wireless communication standard \cite{polar_5G}. 
For improved decoding performance, they are often concatenated to a cyclic redundancy check (CRC) code.
In the remainder of this work, a BCH code with $n=127$, $k=113$ that can correct 2 errors is considered, and addressed as BCH(127,113,2).
Its generator polynomial is $g_\mathcal{C}$=0x7761 defined on GF($2^7$), with a field generator polynomial $g_\mathcal{F}$=0x91.  
The 5G standard $n=128$ $k=105$ polar code concatenated with an 11-bit CRC is used, and labeled PC(128,105)+CRC(11).

\section{LUT-aided error-pattern schedule}\label{sec:LAGRAND}

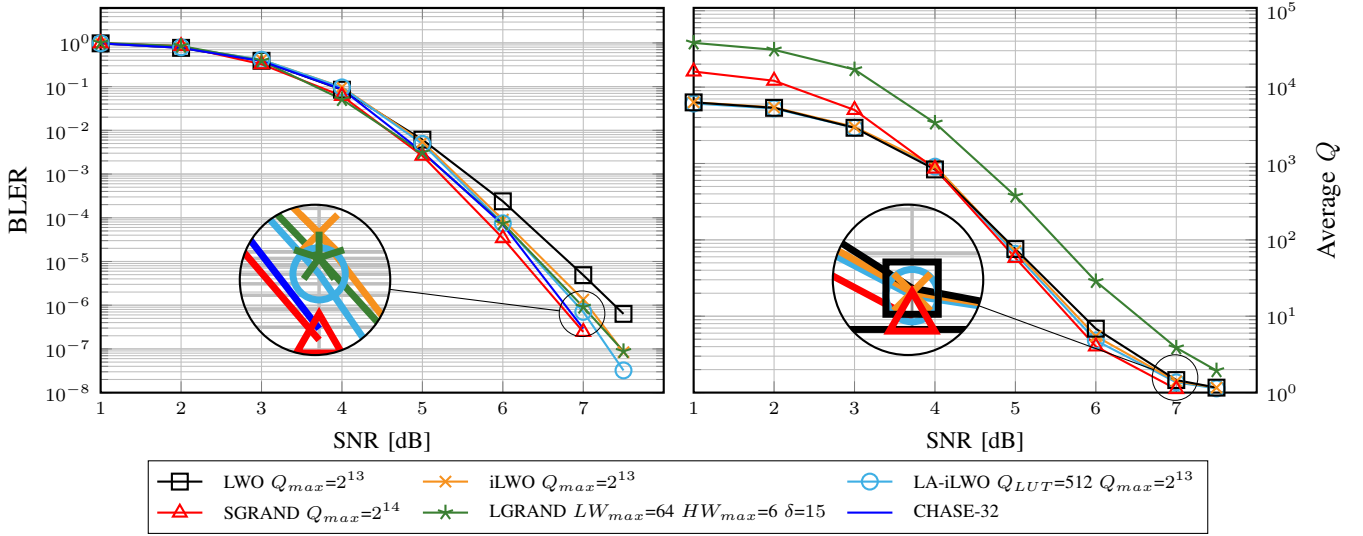
\begin{figure*}[t!]
    \centering
    \begin{minipage}{.5\textwidth}
        \centering
        		  \vspace{5pt}
		  \begin{tikzpicture}[spy using outlines={circle, magnification=3.3, size=2cm, connect spies, transform shape}]
  \pgfplotsset{
    label style = {font=\fontsize{9pt}{7.2}\selectfont},
    tick label style = {font=\fontsize{7pt}{7.2}\selectfont}
  }

\begin{axis}[
	scale = 1,
    ymode=log,
    xlabel={SNR [\text{dB}]}, xlabel style={yshift=0.4em},
    ylabel={BLER}, ylabel style={yshift=-0.75em},
	ylabel near ticks, yticklabel pos=left,
    grid=both,
    ymajorgrids=true,
	yminorgrids=true,
	xmajorgrids=true,
    grid style=solid,
    mark options=solid,
    width=1\columnwidth,
    height=6.7cm,
	xtick={1,2,3,4,5,6,7},
    thick,
        xmin=1,
        xmax=8,
        ymin=1e-8,
		max space between ticks=20,
    mark size=3,
    legend style={
      anchor={center},
      cells={anchor=west},
      mark options=solid,
      column sep= 2mm,
      font=\fontsize{7pt}{7.2}\selectfont,
    },
    legend to name=BLER_ORB_opt_5,
    legend columns=3,
]

\addplot[
    color=black,
    mark=square,
    thick,
    mark size=3,
]
table {
1           0.978      
2           0.771      
3            0.39      
4           0.085      
5       0.0061252      
6     0.000238016      
7    4.87782e-006      
7.5    6.34242e-007    
};
\addlegendentry{LWO $Q_{max}$=$2^{13}$}

\addplot[
    color=BurntOrange,
    mark=x,
    thick,
    mark size=3,
]
table {
1           0.977    
2            0.79    
3           0.412    
4           0.097    
5      0.00544188    
6    9.06201e-005    
7    1.34482e-006    
7.5  8.40837e-008		
};
\addlegendentry{iLWO $Q_{max}$=$2^{13}$}

\addplot[
    color=CornflowerBlue,
    mark=o,
    thick,
    mark size=3,
]
table {
1           0.978 
2           0.795 
3           0.413 
4           0.096 
5      0.00489764 
6    7.48037e-005 
7    7.04311e-007       
7.5  3.22412e-008
};
\addlegendentry{LA-iLWO $Q_{LUT}$=512 $Q_{max}$=$2^{13}$}

\addplot[
    color=red,
    mark=triangle,
    thick,
    mark size=3,
]
table {
1 0.975
2 0.8441
3 0.3251
4 0.0613
5 2.6197e-003
6 3.4066e-005
7 2.4503e-007
};
\addlegendentry{SGRAND $Q_{max}$=$2^{14}$}

\addplot[
    color=OliveGreen,
    mark=star,
    thick,
    mark size=3,
]
table {
1 1
2 0.848
3 0.388
4 0.051
5 0.0031
6 7.4211e-005
7 9.0863e-007
7.5 8.8211e-008
};
\addlegendentry{LGRAND $LW_{max}$=64 $HW_{max}$=6 $\delta$=$15$}

\addplot[
    color=blue,
    thick,
]
table {
1           0.974  
2            0.77  
3           0.384  
4           0.087  
5      0.00320431  
6    6.94443e-005  
7    2.9263e-007
};
\addlegendentry{CHASE-32}

%
%
%
%
%

\spy [black] on (6.40,1.05) in node [left] at (3.85,1.5);

\end{axis}
\end{tikzpicture}
		        \end{minipage}%
    \begin{minipage}{0.5\textwidth}
        \centering
		  \begin{tikzpicture}[spy using outlines={circle, magnification=3.3, size=2cm, connect spies, transform shape}]
  \pgfplotsset{
    label style = {font=\fontsize{9pt}{7.2}\selectfont},
    tick label style = {font=\fontsize{7pt}{7.2}\selectfont}
  }

\begin{axis}[
	scale = 1,
    ymode=log,
    xlabel={SNR [\text{dB}]}, xlabel style={yshift=0.4em},
    ylabel={Average $Q$}, ylabel style={yshift=-0.75em},
	ylabel near ticks, yticklabel pos=right,
    grid=both,
    ymajorgrids=true,
	yminorgrids=true,
    xmajorgrids=true,
    grid style=solid,
    mark options=solid,
    width=1\columnwidth,
    height=6.7cm,
	xtick={1,2,3,4,5,6,7},
    thick,
        xmin=1,
        xmax=8,
        ymin=1,
    mark size=3,
    legend style={
      anchor={center},
      cells={anchor=west},
      mark options=solid,
      column sep= 2mm,
      font=\fontsize{7pt}{7.2}\selectfont,
    },
    legend to name=BLER_ORB_opt_5_comp,
    legend columns=2,
]

\addplot[
    color=CornflowerBlue,
    mark=o,
    thick,
    mark size=3,
]
table {
1    6105.93 
2    5253.54 
3    2932.42 
4     908.76 
5     65.578 
6    4.81502 
7   1.361959    
7.5  1.132818
};
\addlegendentry{LA-iLWO $Q_{LUT}$=512, $Q_{max}$=$2^{13}$}

\addplot[
    color=BurntOrange,
    mark=x,
    thick,
    mark size=3,
]
table {
1      6334.9 
2     5430.81 
3     3014.29 
4     931.253 
5     71.5359 
6     5.47778 
7    1.405464 
7.5  1.161221 	
};
\addlegendentry{iLWO $Q_{max}$=$2^{13}$}

\addplot[
    color=OliveGreen,
    mark=star,
    thick,
    mark size=3,
]
table {
1 38000
2 30800
3 17000
4 3400
5 375
6 28.6
7 3.8654
7.5 1.9433
};
\addlegendentry{LGRAND $LW_{max}$=64 $HW_{max}$=6 $\delta$=$15$}

\addplot[
    color=black,
    mark=square,
    thick,
    mark size=3,
]
table {
1      6313.11   
2      5354.64   
3      2935.77   
4      836.174   
5      75.8634   
6      6.86739   
7     1.458836   
7.5     1.160988 
};
\addlegendentry{LWO $Q_{max}$=$2^{13}$}

\addplot[
    color=red,
    mark=triangle,
    thick,
    mark size=3,
]
table {
1 16000
2 12111
3 5021
4 856
5 58
6 4.02
7 1.11
};
\addlegendentry{SGRAND $Q_{max}$=$2^{14}$}

\addplot[
    color=blue,
    thick,
]
table {
1 0
2 0
3 0
4 0
5 0
6 0
7 0
};
\addlegendentry{CHASE-32}

%
%
%
%
%

\spy [black] on (6.40,0.2) in node [left] at (3.85,1.5);
\end{axis}
\end{tikzpicture}
   \end{minipage}
    \ref{BLER_ORB_opt_5}
    \caption{BLER and average $Q$ for BCH(127,113,2).}
    \label{fig:BCH}
\end{figure*}

In \cite{iLWO}, the improved logistic weight order (iLWO) schedule was proposed.
It is an ordering of error patterns for ORBGRAND decoding that computes the weight of each error pattern $\mathbf{e}$ as follows: 
\begin{equation}
iLW(\mathbf{e}) = \sum_{i=0}^{HW-1} (i+1)\cdot(v_i+1)~,
\end{equation}
where $\mathbf{v}$ has been defined for (\ref{eq:LW}).
The iLWO resulting from the above $iLW$ favors patterns with low $HW$ and potentially high $LW$ over patterns with low $LW$ but high $HW$. 
It follows the observation that at high enough signal-to-noise ratio (SNR), and thus low enough block error rate (BLER), high-$HW$ patterns are less likely to lead to successful decoding, but LWO can schedule them with high priority.
iLWO has been shown to yield substantial gains in error-correction performance with respect to LWO, it can be generated algorithmically, and it is parallelizable.
However, it is an approximation of the sequence of error patterns that can be observed empirically on an AWGN channel, and as such it is suboptimal.
To further improve the error-correction performance of ORBGRAND and reduce its average number of queries $Q$ given a maximum $Q_{max}$, the look-up-table-aided (LUT-aided) error-pattern schedule is proposed here.
It first attempts the decoding with $Q_{LUT}$ empirically-observed error patterns stored in order of likelihood of occurrence, while the remaining $Q_{max}-Q_{LUT}$ patterns are generated according to the preferred schedule, excluding the intersection with the set of the already selected $Q_{LUT}$ patterns.
Since iLWO currently yields the best error-correction performance and lowest average $Q$ among the error-pattern schedules for ORBGRAND in literature, in the remainder of this work the LUT-aided iLWO is considered, and it is labeled LA-iLWO. 
Nevertheless, LUT-aided decoding is a general approach that can be paired with any other schedule if so desired, for example LWO \cite{ORBGRAND_first} or the $HW$-based schedule of GRAND \cite{GRAND_first}. 
In the same way, LUT-aided schedules can be easily implemented in any practical ORBGRAND decoder \cite{ORBARCH_J}, improving error-correction performance and speed.

\begin{figure*}[t!]
    \centering
    \begin{minipage}{.5\textwidth}
        \centering
        		   \vspace{5pt}
		  \begin{tikzpicture}[spy using outlines={circle, magnification=3.3, size=2cm, connect spies, transform shape}]
  \pgfplotsset{
    label style = {font=\fontsize{9pt}{7.2}\selectfont},
    tick label style = {font=\fontsize{7pt}{7.2}\selectfont}
  }

\begin{axis}[
	scale = 1,
    ymode=log,
    xlabel={SNR [\text{dB}]}, xlabel style={yshift=0.4em},
    ylabel={BLER}, ylabel style={yshift=-0.75em},
	ylabel near ticks, yticklabel pos=left,
    grid=both,
    ymajorgrids=true,
	yminorgrids=true,
	xmajorgrids=true,
    grid style=solid,
    mark options=solid,
    width=1\columnwidth,
    height=6.7cm,
	xtick={1,2,3,4,5,6,7},
    thick,
        xmin=1,
        xmax=8,
        ymin=1e-8,
		max space between ticks=20,
    mark size=3,
    legend style={
      anchor={center},
      cells={anchor=west},
      mark options=solid,
      column sep= 2mm,
      font=\fontsize{7pt}{7.2}\selectfont,
    },
    legend to name=BLER_polar,
    legend columns=4,
]

\addplot[
    color=black,
    mark=square,
    thick,
	dashed,
    mark size=3,
]
table {
1           0.962   
2           0.777   
3           0.378   
4       0.0861326   
5      0.00476054   
6     0.000105172   
7    7.46962e-007   
7.5  5.0000e-008
};
\addlegendentry{LWO $Q_{max}$=$2^{17}$}

\addplot[
    color=BurntOrange,
    mark=x,
    thick,
	dashed,
    mark size=3,
]
table {
1           0.962     
2           0.777     
3           0.378     
4       0.0861326     
5      0.00476054     
6     0.000102933     
7    6.86494e-007     
7.5  4.6214e-008
};
\addlegendentry{iLWO $Q_{max}$=$2^{17}$}

\addplot[
    color=black,
    mark=square,
    thick,
    mark size=3,
]
table {
1           0.984     
2           0.862     
3           0.514     
4           0.144     
5       0.0106963     
6     0.000435908     
7    1.06844e-005     
7.5    1.49232e-006   
};
\addlegendentry{LWO $Q_{max}$=$2^{13}$}

\addplot[
    color=BurntOrange,
    mark=x,
    thick,
    mark size=3,
]
table {
1           0.988   
2           0.865   
3           0.537   
4           0.161   
5        0.011157   
6      0.00021153      
7      1.167e-006   
7.5    5.306e-008     
};
\addlegendentry{iLWO $Q_{max}$=$2^{13}$}

\addplot[
    color=CornflowerBlue,
    mark=o,
    thick,
    mark size=3,
]
table {
1           0.988   
2            0.87   
3           0.544   
4           0.164   
5       0.0117316   
6      0.000192709 
7    1.14125e-006   
7.5  5.1586e-008
};
\addlegendentry{LA-iLWO $Q_{LUT}$=512 $Q_{max}$=$2^{13}$}

\addplot[
    color=red,
    mark=triangle,
    thick,
    mark size=3,
]
table {
1 0.961
2 0.454
3 0.0713
4 0.0039
5 0.00013
6 2.9877e-006
7 9.5065e-008
};
\addlegendentry{SGRAND $Q_{max}$=$2^{23}$}

\addplot[
    color=CornflowerBlue,
    mark=o,
    thick,
	dashed,
    mark size=3,
]
table {
1           0.962 
2           0.777 
3           0.378 
4       0.0861326 
5      0.00476054 
6     0.000105096 
7    6.51953e-007
7.5  4.52708e-008
};
\addlegendentry{LA-iLWO $Q_{LUT}$=512 $Q_{max}$=$2^{17}$}

\addplot[
    color=blue,
    thick,
]
table {
1            0.97    
2            0.79    
3           0.432    
4       0.0814996    
5      0.00412167    
6    4.28089e-005    
7	 1.17285e-007
};
\addlegendentry{SCL-8}

\spy [black] on (6.40,1.14) in node [left] at (3.85,1.5);

\end{axis}
\end{tikzpicture}
		        \end{minipage}%
    \begin{minipage}{0.5\textwidth}
        \centering
         \vspace{5pt}
		  \begin{tikzpicture}[spy using outlines={circle, magnification=3.4, size=1.8cm, connect spies, transform shape}]
  \pgfplotsset{
    label style = {font=\fontsize{9pt}{7.2}\selectfont},
    tick label style = {font=\fontsize{7pt}{7.2}\selectfont}
  }

\begin{axis}[
	scale = 1,
    ymode=log,
    xlabel={SNR [\text{dB}]}, xlabel style={yshift=0.4em},
    ylabel={Average $Q$}, ylabel style={yshift=-0.75em},
	ylabel near ticks, yticklabel pos=right,
    grid=both,
    ymajorgrids=true,
	yminorgrids=true,
    xmajorgrids=true,
    grid style=solid,
    mark options=solid,
    width=1\columnwidth,
    height=6.7cm,
	xtick={1,2,3,4,5,6,7},
    thick,
        xmin=1,
        xmax=8,
        ymin=1,
    mark size=3,
    legend style={
      anchor={center},
      cells={anchor=west},
      mark options=solid,
      column sep= 2mm,
      font=\fontsize{7pt}{7.2}\selectfont,
    },
    legend to name=BLER_polar_comp,
    legend columns=2,
]

\addplot[
    color=CornflowerBlue,
    mark=o,
    thick,
	dashed,
    mark size=3,
]
table {
1     125930
2     103725
3    53390.7
4    12845.5
5    872.046
6    25.9411
7    1.84828
7.5     1.292
};
\addlegendentry{LA-iLWO $Q_{LUT}$=512 $Q_{max}$=$2^{17}$}

\addplot[
    color=BurntOrange,
    mark=x,
    thick,
	dashed,
    mark size=3,
]
table {
1   126337
2   104044
3    53526
4  12877.1
5  879.307
6  27.2416
7  1.96216	
7.5     1.358
};
\addlegendentry{iLWO $Q_{max}$=$2^{17}$}

\addplot[
    color=CornflowerBlue,
    mark=o,
    thick,
    mark size=3,
]
table {
1     7692.09
2     6990.72
3     4876.89
4     1711.27
5     227.467
6     12.9051
7     1.77321
7.5     1.191
};
\addlegendentry{LA-iLWO $Q_{LUT}$=512 $Q_{max}$=$2^{13}$}

\addplot[
    color=BurntOrange,
    mark=x,
    thick,
    mark size=3,
]
table {
1   8116.08
2   7359.05
3   5097.26
4   1782.39
5   236.041
6   14.4367
7   1.88744
7.5     1.263
};
\addlegendentry{iLWO $Q_{max}$=$2^{13}$}

\addplot[
    color=black,
    mark=square,
    thick,
    mark size=3,
]
table {
1     8102.63
2     7308.87
3     4878.02
4     1649.33
5     215.879
6     17.0816
7     2.10838
7.5     1.34043
};
\addlegendentry{LWO $Q_{max}$=$2^{13}$}

\addplot[
    color=red,
    mark=triangle,
    thick,
    mark size=3,
]
table {
1 8088608
2 3305221
3 510218
4 29840
5 482
6 9.72
7 1.69
};
\addlegendentry{SGRAND $Q_{max}$=$2^{23}$}

\addplot[
    color=black,
    mark=square,
    thick,
	dashed,
    mark size=3,
]
table {
1      126180
2      103475
3       52848
4     12451.6
5     861.643
6     33.4168
7     2.31221
7.5     1.68211
};
\addlegendentry{LWO $Q_{max}$=$2^{17}$}

\addplot[
    color=blue,
    thick,
]
table {
7	 0
};
\addlegendentry{SCL-8}

\addplot[
    color=OliveGreen,
    mark=star,
    thick,
    mark size=3,
]
table {

};
\addlegendentry{LGRAND $LW_{max}$=64 $HW_{max}$=6 $\delta=15$}

\spy [black] on (6.40,0.2) in node [left] at (3.85,1.0);

\end{axis}
\end{tikzpicture}
   \end{minipage}
    \ref{BLER_polar}
    \caption{BLER and average $Q$ for PC(128,105)+CRC(11).}
    \label{fig:polar}
\end{figure*}
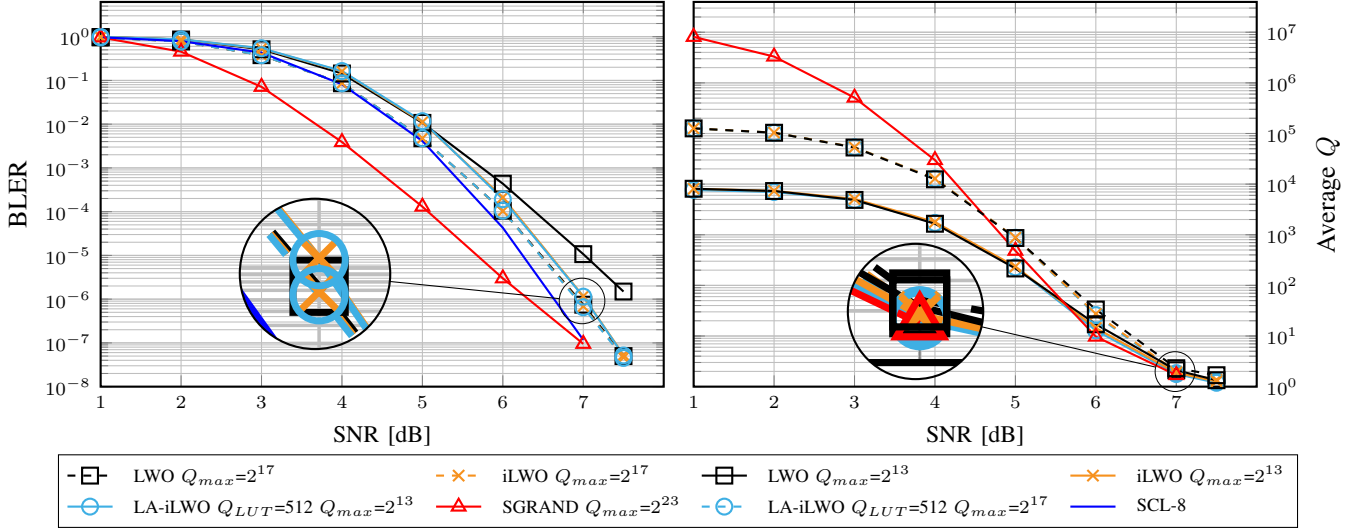

Figure \ref{fig:BCH} plots the BLER and average number of queries $Q$ for a BCH(127,113,2) code, decoded with various GRAND-based decoders, and with the soft-input type-II Chase decoder \cite{CHASE} with 32 test patterns. 
The latter is a common choice for high-performance decoding of BCH codes.
Simulations consider BPSK modulation over an AWGN channel.
SGRAND \cite{soft_GRAND} with $Q_{max}=2^{14}$ achieves ML performance with this code at the observed SNRs, and it is plotted as a baseline. 
It can be seen that Chase-32 follows SGRAND closely in the considered SNR range.
The blue curve with circular markers portrays the performance of ORBGRAND with the novel LA-iLWO schedule, $Q_{max}=2^{13}$, and $Q_{LUT}=512$. 
To select the $Q_{LUT}$ error patterns, the transmission of 128-bit vectors was simulated, and $10^9$ error events at SNR$=7$ dB were observed, totaling around 2000 different error patterns on the sorted hard-decided vector $\pi({\rm HD}(\mathbf{y}))$.
The 512 ones occurring most frequently account for more than $99.85\%$ of the total; their frequency of occurrence is monotonically descending, and each point of the curve counts more than 100 observations.
The selected $Q_{LUT}$ patterns vary with the target SNR, albeit very gradually, and are expected to change in presence of different channel models.
The remaining $Q_{max}-Q_{LUT}=7680$ patterns are created according to iLWO, excluding the patterns already included in the first $Q_{LUT}$. 
LA-iLWO is shown to yield a 0.15 dB gap from ML at BLER$=10^{-6}$, while outperforming both iLWO (0.05 dB gain) and LWO (0.4 dB gain) with the same $Q_{max}$.
At the same time, it matches the performance of list-GRAND (LGRAND, \cite{LGRAND}), that follows LWO and uses approximately $5\times$ $Q_{max}$, together with a more complex algorithm. 
iLWO is shown to also closely follow LGRAND as well, regardless of its lower complexity.
At BLER$=10^{-7}$, the gain over LWO, iLWO and LGRAND increases.
The average $Q$ of LWO, iLWO and LA-iLWO are very close, with iLWO having lower $Q$ than LWO, and LA-iLWO outperforming both at high enough SNR, whereas the high $Q_{max}$ of LGRAND leads to substantially higher $Q$.

Figure \ref{fig:polar} shows BLER and average $Q$ for a polar code PC(128,105) concatenated with an 11-bit cyclic redundancy check (CRC), both taken from the 5G standard \cite{polar_5G}.
SGRAND, again plotted as a reference, requires $Q_{max}=2^{23}$ to achieve ML performance. 
Considering $Q_{max}=2^{13}$, LA-iLWO has a 0.65 dB gap from ML performance at BLER$=10^{-6}$. 
LA-iLWO and iLWO have almost the same performance, yielding a gain of more than 0.5 dB with respect to LWO; all three ORBGRAND schedules have very similar average average $Q$, with that of LA-iLWO being the lower of the three at high SNR.
The gap between LA-iLWO and SGRAND reduces to 0.35 dB at BLER$=10^{-7}$.
Increasing $Q_{max}$ to $2^{17}$ makes LA-iLWO, iLWO, and LWO yield almost the same performance, while their average $Q$ maintain the same trend observed for $Q_{max}=2^{13}$. 
This is due to the fact that as $Q_{max}$ increases, the first $Q_{LUT}$ patterns of LA-iLWO are eventually scheduled by iLWO as well, leading to similar BLER at a higher average $Q$. 
Increasing $Q_{max}$ allows also the suboptimal LWO to schedule patterns that were given a higher priority by both LA-iLWO and iLWO.
The performance of the widespread successive cancellation list decoder (SCL, \cite{list_decoding}) with a list size of 8, often considered the state-of-the-art decoder for 5G polar codes, is plotted as well.
It can be seen that it matches the performance of LA-iLWO at lower SNR, while quickly converging towards that of SGRAND at higher SNR.
It is worth noting that the $Q_{LUT}$ patterns used by LA-iLWO to decode a polar code in Figure \ref{fig:polar} are the same ones used to decode a BCH code in Figure \ref{fig:BCH}, since the error patterns depend on the code length but are independent from the encoding used, if any.

The maximum number of queries $Q_{max}$ is one of the main factors in determining the worst case latency of a GRAND-based decoder. 
By considering high-performance error-pattern schedules like LA-iLWO and iLWO, that schedule error patterns with high likelihood of occurrence earlier than LWO, ORBGRAND can achieve performance close to ML with a relatively small $Q_{max}$ and low average $Q$.
These schedules are thus well suited for practical implementation, as they can reduce latency, increase throughput and reduce power consumption of hardware decoders.

\section{Decoder Architecture} \label{sec:arch}

\begin{figure*}[t!]
	\centering
	\includegraphics[width=\linewidth]{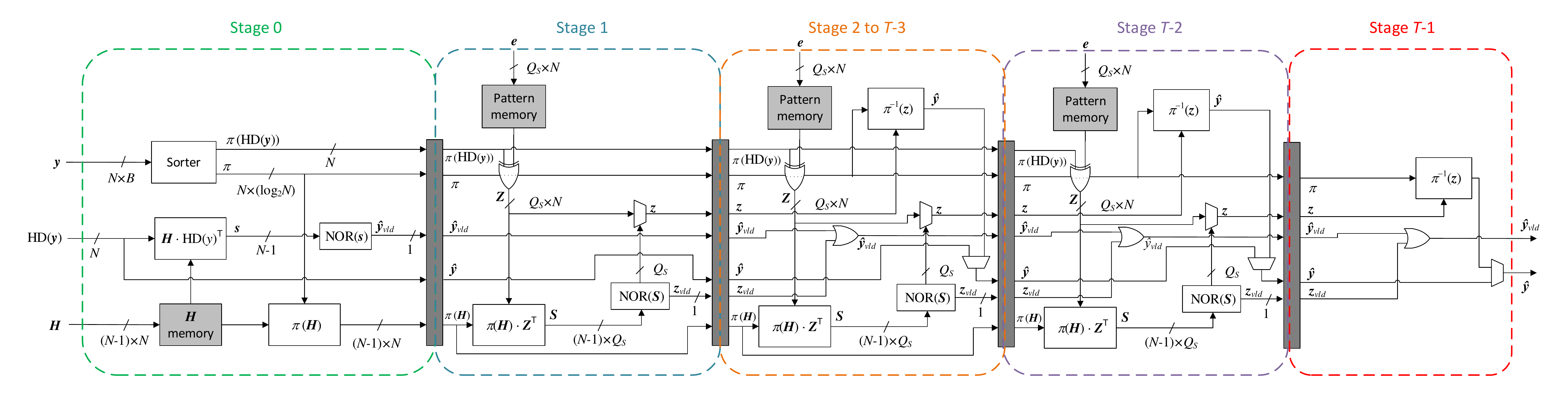}
	\caption{Proposed ORBGRAND decoder architecture.}
	\label{fig:DEC}
\end{figure*}

GRAND-based decoder architectures in literature rely on the reuse of hardware resources for successive decoding attempts. 
This approach yields very short average latency and very high average throughput, but incurs dismal figures of merit in the worst case scenario. 
This drawback limits their appeal in practical applications where strict latency constraints are present.
In this Section, a decoder architecture for ORBGRAND-based decoding that has a very high fixed throughput and a very short fixed latency is presented. 
In the spirit of GRAND-based algorithms, the decoder is code-agnostic: it can decode any type of binary linear code with code length $\le N$ and with code rate $\ge 1/N$.
The decoder is based on a highly-parallel, pipelined, feed-forward design; while the architecture is not tied to a particular error-pattern schedule, the decoder leverages the improved performance of LA-iLWO to limit $Q_{max}$. 
Figure \ref{fig:DEC} shows the proposed decoder architecture: it is divided into $T$ stages separated by registers, that are represented by gray rectangles.

\subsection{Stage 0}

The LLR vector $\mathbf{y}$ is input in stage $0$, each of the $N$ LLRs represented with sign and magnitude over $B$ bits. 
It is sorted by a bitonic sorter that uses $B-1$ bits for internal comparisons, i.e. the magnitude of each LLR, while also forwarding the sign bit of each LLR and its $\lg2_2N$-bit index. 
Consequently, each of the $\lg2_2N$ pipeline stages internal to the sorter are constituted of $(B+\lg2_2N)\times N$ registers.
Note that assuming BPSK modulation that maps $0\rightarrow+1$ and $1\rightarrow-1$, ${\rm HD}(\mathbf{y})$ is equal to the vector of sign bits of $\mathbf{y}$. 
The outputs of the sorter are the permuted vector of sign bits $\pi({\rm HD}(\mathbf{y}))$ and the $N\times \lg2_2N$ sorted vector of indices $\pi$, that can be used as a permutation vector to recover the natural order corrected vector $\mathbf{\hat{y}}$ by one of the following stages.
If a code with code length $n<N$ is to be decoded, the $n$ lowest indices of $\mathbf{y}$ have to contain the valid LLRs. 
The magnitudes of the unused $N-n$ LLRs need to be set to $2^{B-1}-1$, so that the sorter does not permute them, and their sign bits set to 0.

\begin{figure}[t!]
	\centering
	\hspace{-15pt}
	\includegraphics[width=1.05\linewidth]{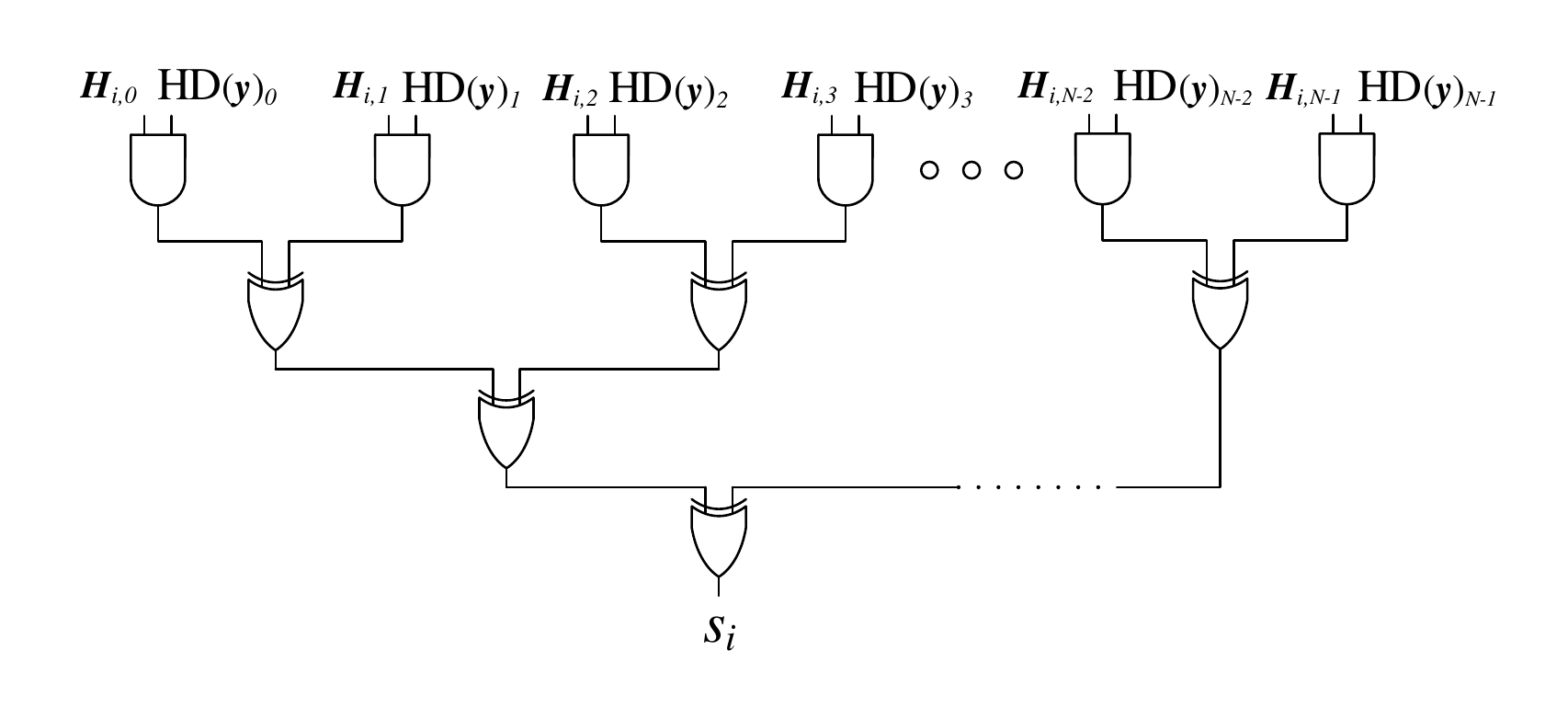}
	\caption{Syndrome calculation circuit for stage 0.}
	\label{fig:HmatMult}
\end{figure}

Parallel to the sorter, the $N$-bit vector containing the sign of each LLR ${\rm HD}(\mathbf{y})$ is multiplied by the parity check matrix of the code $\mathbf{H}$. 
The $\mathbf{H}$ matrix is stored in a dedicated register matrix of dimension $(N-1) \times N$ ($\mathbf{H}$ memory in Figure \ref{fig:DEC}).
Since $\mathbf{H}$ changes every time a new code is used, the $\mathbf{H}$ memory needs to be programmable; to allow that, an $N$-bit input is used to overwrite the contents of the $\mathbf{H}$ memory row-by-row after a code change.
The $\mathbf{H} \cdot {\rm HD}(\mathbf{y})^{\rm T}$ product is performed for each row of $\mathbf{H}$ in parallel, as shown in Figure \ref{fig:HmatMult}: to compute the result of parity check $i$, bit ${\rm HD}(\mathbf{y})_j$ with $0\le j < N$ is ANDed with $\mathbf{H}_{i,j}$, and the $N$ resulting bits are XORed together to obtain the syndrome bit $s_i$. 
In case $s_i=0$, $0\le i < N-1$ then ${\rm HD}(\mathbf{y})$ is already a valid codeword: it is forwarded to the following stages as the valid candidate in natural order $\mathbf{\hat{y}}$, and the $\mathbf{\hat{y}}_{vld}$ flag is set to 1 by NORing all bits of $\mathbf{s}$.
The $N-1$ rows of $\mathbf{H}$ allow for code rates as low as $1/N$, but in case a lower range of rates is foreseen, the matrix dimensions become $(N(1-R_{min}))\times N$, where $R_{min}$ is the minimum code rate allowed by the system.
Given a code rate $r\ge R_{min}$ and a code length $n\le N$, the $N-n$ rightmost columns and the $N(1-(r-R_{min}))$ lowermost rows of $\mathbf{H}$ need to be set to zero, so that the unused symbols are ignored in all the parity checks, and that all syndromes relative to unused parity checks are zero. 

The columns of $\mathbf{H}$ are also permuted according to $\pi$, the resulting $\pi(\mathbf{H})$ being forwarded to the next stages, where it is used to verify if one of the vectors obtained through ORBGRAND in permuted order is a valid codeword.

In case ${\rm HD}(\mathbf{y})$ is a valid codeword, the decoding process has finished, and only $\mathbf{\hat{y}}={\rm HD}(\mathbf{y})$ and $\mathbf{\hat{y}}_{vld}=1$ need to be forwarded to the following stages: consequently, all other registers between Stage 0 and 1 are not updated, reducing the dynamic power consumption.
Moreover, as soon as $\mathbf{\hat{y}}_{vld}=1$ is available, the sorting operation can be stopped, allowing to avoid the update of the last $\lg2_2N-2$ pipeline stages internal to the sorter and further save power. 
Unlike architectures like \cite{ORBARCH_J}, where successful decoding with $Q<Q_{max}$ allows to increase the average throughput, the proposed architecture uses the variable $Q$ as a means to limit switching activity and reduce power consumption. 

\subsubsection{Sorter pruning}

\begin{figure}[t!]
	\centering
	\includegraphics[width=1.05\linewidth]{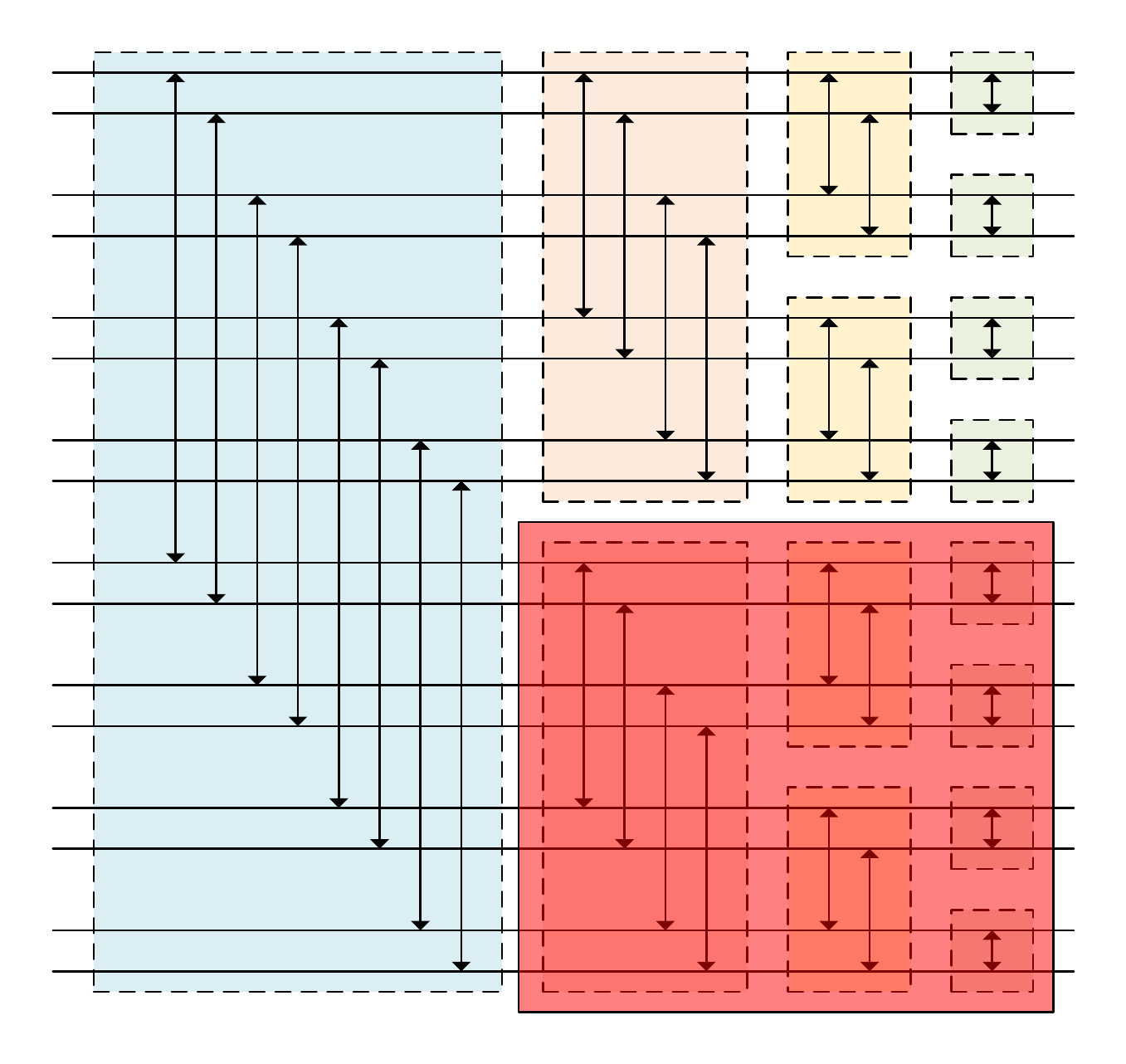}
	\caption{Last stage of an $N=16$ bitonic sorter.}
	\label{fig:bitsort}
\end{figure}

Depending on $Q_{max}$, $N$, and the chosen pattern schedule, the bitonic sorter can be pruned without loss of error-correction performance and at undetectable cost in average $Q$. 
Let us consider the parameters used in Section \ref{sec:impl} for the decoder implementation (LA-iLWO, $Q_{max}=2^{13}$, $Q_{LUT}=512$, $N=128$): with these conditions, while computing $\pi({\rm HD}(\mathbf{y})) \oplus \mathbf{e}$, the last $N/2$ entries of $\pi({\rm HD}(\mathbf{y}))$ are flipped only by $\mathbf{e}$ with $HW=1$. 
In fact, the scheduled error patterns with $HW>1$ involve only the first $N/2$ positions of $\pi({\rm HD}(\mathbf{y}))$.
Consequently, any shuffling among the last $N/2$ positions will not have any impact on the error-correction capability of the decoder, at most changing the number of average queries $Q$.
The structure of a bitonic sorter foresees $\lg2_2N$ stages, each one with an additional set of compare-and-swap units \cite{bitsort}.
The last stage, shown in Figure \ref{fig:bitsort} for $N=16$, performs $\lg2_2N$ consecutive sets of  compare-and-swap. 
After the first set, the $N/2$ most reliable LLRs have been separated from the $N/2$ least reliable ones: the $\lg2_2N-1$ sets of compare-and-swap involving only the $N/2$ least reliable LLRs, highlighted in red in the figure, can be removed, as they would implement only an internal permutation.
No change in in average $Q$ could be detected via simulation using the pruned sorter.

\subsection{Stage 1}

After the decoder stage 0 has performed preliminary operations (checking if $\mathbf{y}$ represents a valid codeword, sorting the LLRs and providing $\pi$), stage 1 is where the core of GRAND-based decoding begins.

The LUT-aided schedule described in Section \ref{sec:LAGRAND} requires the offline identification of the $Q_{LUT}$ initial patterns, while the remaining $Q_{max} - Q_{LUT}$ can be generated on-the-fly via LWO, iLWO, or any other schedule. 
Nevertheless, the pipelined nature of the proposed architecture demands that all $Q_{max}$ error patterns are available at the same time, albeit in different pipeline stages.
Consequently, all $Q_{max}$ patterns are pre-generated and stored in dedicated memories; the $Q_{max}$ patterns are divided among $Q_{max}/Q_S$ decoder stages, with each stage attempting $Q_S$ error patterns, starting with stage 1.
A pattern memory stores $Q_S$ binary vectors of $N$ bits; each vector is a pattern $\mathbf{e}$ with a 1 at every position of the sorted hard-decision vector $\pi({\rm HD}(\mathbf{y}))$ to be flipped. 
Much like the $\mathbf{H}$ memory in stage 0, the pattern memory is a $Q_S \times N$ matrix of registers that can be programmed one row at a time.
The storage of precomputed patterns enables the application of pattern schedules that cannot be easily generated algorithmically, like the first $Q_{LUT}$ patterns of LA-iLWO, and removes the need to constrain the $HW$ and $LW$ of patterns \cite{ORBARCH_J}.
As the implementation cost of the pattern memories can grow quickly with $Q_{max}$, this solution suits well high-performance schedules like LA-iLWO, that rely on smaller $Q_{max}$.

All $Q_S$ patterns are XORed in parallel to $\pi({\rm HD}(\mathbf{y}))$, returning the $Q_S \times N$ matrix $\mathbf{Z}$, where $\mathbf{z}_i = \pi({\rm HD}(\mathbf{y})) \oplus \mathbf{e}_i$, $0\le i < Q_S$. 
Multiplying each vector $\mathbf{z}_i$ to the permuted $\mathbf{H}$ matrix $\pi({\mathbf{H}})$, the syndrome matrix $\mathbf{S}$ is obtained, each column $\mathbf{s}_i$ the syndrome of $\mathbf{z}_i$.
If one of the syndromes is the all-zero vector, then one of the rows of $\mathbf{Z}$ is a valid codeword, and the $\mathbf{z}_{vld}$ flag is set to 1 and forwarded to the next stage.
Since the error patterns are ordered but they are applied in parallel, it is possible that more than one syndrome is zero; a priority selector forwards to the next stage the $\mathbf{z}_i$ corresponding to the zero syndrome with the highest priority.

In case $\mathbf{z}_{vld}$ is risen, the pipeline registers for $\mathbf{\hat{y}}$, $\pi({\rm HD}(\mathbf{y}))$, $\pi(\mathbf{H})$ are not enabled, as a valid codeword was just found and the following stages do not need to continue the decoding.
If $\mathbf{\hat{y}}_{vld}=1$ at the input of stage 1, then a valid codeword was found at stage 0, and also the pipeline registers corresponding to $\mathbf{z}$ and $\pi$ are disabled, while $\mathbf{\hat{y}}$ is activated.
In case no valid codeword was found at stage 0 and 1, then $\mathbf{\hat{y}}_{vld}=\mathbf{z}_{vld}=0$, and the $\mathbf{z}$ and $\mathbf{\hat{y}}$ registers are disabled, while $\pi({\rm HD}(\mathbf{y}))$, $\pi(\mathbf{H})$, and $\pi$ are enabled.

\ctable[
	caption={Proposed decoder implementation results versus \cite{GRANDchip, BCHDEC2}.},
	star,
	label={tab:BCH}, 	
    ]{l|rrr|r|r}{
    \tnote[a]{GRAND algorithm}
    \tnote[b]{$n$=\{128,127\} only}
    \tnote[c]{Modified type-II Chase algorithm with 4 test patterns}
    \tnote[d]{Single code decoder for BCH(63,51,2)}
    }{
    	\toprule
	&  \multicolumn{3}{c|}{This work} & \cite{GRANDchip}\tmark[a] & \cite{BCHDEC2}\tmark[c] \\
	              &  {\bf A$_{\rm BCH}$} & {\bf B} & {\bf C} &  &\\
\midrule
		Result type & \multicolumn{3}{c|}{Synthesis} & Fabricated & Synthesis \\ 
		Technology [nm]  &  7 & 7 & 7 & 40 & 90\\
		Supply [V]  &  0.5 & 0.5 & 0.5 & 1.1 & 1.0\\
		$N$ 	 	&  128 				&	128       &	128    &  128\tmark[b] & 63\tmark[d]\\
		$R_{min}$   &  1/128 			&	0.656      &	0.656    &  0.656 &  51/63\\
		$r$         &  113/127 			&	113/127      &	113/127    &  113/127 & 51/63\\
		$Q_{max}$	&  $2^{13}$			&	$2^{13}$  &	$2^{13}$  & $1.33\cdot2^{18}$ & -\\
		$Q_S$		&  512 				&	512       & 256  &  - & -\\
		$T$			&  18				&   18		  & 34 	 &  - & -\\	
		$Q_{LUT}$	&  512				&	512       & 512  &  - & -\\
		$B$         &  8				&  8		&  8	&  - & NA\\
	\midrule
 		$f$ [MHz]   &  616				&	616       &  701  & 68 & 250\\
		Area [mm$^2$] & 5.16 			&   3.70	& 4.05 & 0.83 & 0.17\\
		Area @ 20 nm [mm$^2$] & 18.58			&   13.32	& 14.58  & 0.26 & 0.013 \\
	   \midrule
		B.C. $\mathcal{L}$  ~\multirow{2}{*}{[cc] - [ns]}     & \multirow{2}{*}{25 - 40.58} 	&	\multirow{2}{*}{25 - 40.58} & \multirow{2}{*}{41 - 58.49}  & 71 - 1044.12 &  \multirow{2}{*}{18 - 72}\\
		W.C. $\mathcal{L}$      & 	&	 &   & 29500 - 433823.53 &\\
    	\midrule
		B.C. $\mathcal{T}$  ~\multirow{2}{*}{[$\frac{\rm bits}{\rm cycle}$] - [Gb/s]}   & \multirow{2}{*}{113 - 69.61}     & \multirow{2}{*}{113 - 69.61}   & \multirow{2}{*}{113 - 79.21} & 1.59 - 0.108 &  \multirow{2}{*}{5.7 - 1.417}\\
		W.C. $\mathcal{T}$   &    &  && 0.004 - $260\cdot 10^{-6}$ &\\
    	\midrule
		Area eff. [Gbps/mm$^2$] & 13.49     &  18.81  & 19.56 & 0.13 & 8.34\\	
		Area eff. @ 20 nm [Gbps/mm$^2$] & 3.75     &  3.05  & 5.23 & 0.42 & 109\\	
		Power [mW]  			& 277.66  & 196.67	&204.06  & 3.75 & NA\\
		Energy/bit [pJ/bit] & 3.99	& 2.83 & 2.58  & 30.6 & NA\\ 
		\bottomrule
}

\subsection{Stage 2 to $T$-3}

The $T$-4 stages following stage 1 all have the same functionality, differing only by the contents of the pattern memory.
Beside all of stage 1 operations, these stages also perform the inverse permutation of $\mathbf{z}$ to obtain the natural order candidate codeword $\mathbf{\hat{y}}$.
If stage $t$ receives $\mathbf{z}_{vld}=1$ as an input, a valid codeword in permuted order was identified at stage $t-1$. 
The permutation $\pi$ is then used to perform $\mathbf{\hat{y}}=\pi^{-1}(\mathbf{z})$ through $N$ $N$-to-1 demultiplexers, setting $\mathbf{\hat{y}}_{vld}=1$ and forwarding only $\mathbf{\hat{y}}$ to the following stages, while the other pipeline registers are not enabled.

\subsection{Stage $T$-2 and $T$-1}

Stage $T$-2 is the last one that includes a pattern memory, and thus the last where error correction can be attempted. 
As such, it maintains the functionality of the previous stage, but neither $\pi({\rm HD}(\mathbf{y}))$ nor $\pi(\mathbf{H})$ can be forwarded to the last stage.

The last stage can only perform $\mathbf{\hat{y}}=\pi^{-1}(\mathbf{z})$ if it receives $\mathbf{z}_{vld}=1$, otherwise it outputs the received $\mathbf{\hat{y}}$ and $\mathbf{\hat{y}}_{vld}$.

\subsection{Latency and throughput}

Given that the number of decoder stages $T$ is equal to $Q_{max}/Q_S+2$, the decoding latency of the proposed decoder architecture in clock cycles can be computed as 
\begin{equation}
\mathcal{L}=\frac{Q_{max}}{Q_S}+2+\log_2N~,
\end{equation}
where $\lg2_2N$ is the latency introduced by the sorter in stage 0.
The information throughput in bit/s is instead computed as 
\begin{equation}
\mathcal{T}=nrf~,
\end{equation}
where $n\le N$ is the code length, $r\ge R_{min}$ is the code rate, and $f$ is the clock frequency at which the decoder is working.
Unlike decoder architectures in literature, neither $\mathcal{L}$ nor $\mathcal{T}$ depend on the average $Q$.

\section{Implementation}\label{sec:impl}

The decoder architecture proposed in Section \ref{sec:arch} has been described in Verilog HDL and synthesized in TSMC 7 nm FinFET technology, using a typical corner; implementation results are reported in Table \ref{tab:BCH}-\ref{tab:polar} along with state of the art decoders.
The area and area efficiency of all solutions listed in these Tables have also been scaled to the 20 nm technology node using the scaling factors provided in \cite{scaling}. 
The complexity of the designs and the distance between the native technology nodes involved makes it virtually impossible to have a fair comparison, and the scaled results should be considered as no more than order-of-magnitude approximations.

Table \ref{tab:BCH} showcases the implementation results for different combinations of decoder parameters, for a maximum code length $N=128$ and $Q_{max}=2^{13}$.
Power estimations have been obtained annotating the decoder switching activity using realistic test vectors provided by the simulator: the reported figures consider a BCH(127,113,2) code decoded with ORBGRAND and the LA-iLWO schedule with $Q_{LUT}=512$ patterns observed at SNR$=7$ dB (BLER$=10^{-6}$). 
Nevertheless, the power is estimated at a target bit error rate of $10^{-5}$, the same conditions as \cite{GRANDchip}, whose results are included in the Table.
Implementation {\bf A$_{\rm BCH}$} allows for the highest degree of code rate flexibility by instantiating an $\mathbf{H}$ memory of size $(N-1)\times N$. 
The high degree of internal parallelism inferred by $Q_S=512$ leads to a very short latency of 25 clock cycles, however limiting the achievable frequency to $f=616$ MHz, with an area occupation of 5.16 mm$^2$. 
The information throughput for the considered code is of 69.61 Gb/s.
Since the proposed architecture guarantees fixed latency and throughput, there is no difference between best case (B.C.) and worst case (W.C.) $\mathcal{L}$ and $\mathcal{T}$.
Power consumption at the observed SNR is 277.66 mW, dominated by leakage power.
In fact, more than 54\% of received vectors are valid codewords, thus requiring only partial activation of stage 0, while the vast majority of the remaining ones can be corrected in stage 1, with the activation rate of stage 2 being $\approx 0.3\%$. 
Thus, the switching activity in stages 2 to 17 is virtually null, greatly limiting the dynamic power of the decoder.

Implementation {\bf B} maintains the same decoder parameters as {\bf A$_{\rm BCH}$} but reduces the size of the $\mathbf{H}$ memory, imposing $R_{min}=0.656$, to enable a fairer comparison with \cite{GRANDchip}. 
This change of parameter is reflected on the size of all pipeline stages, on the complexity of the syndrome calculation circuit, and on the generation of $\mathbf{\hat{y}}_{vld}$ (stage 0) and $\mathbf{z}_{vld}$ (stage 1 to $T$-2); the total area occupation is reduced to 3.70 mm$^2$.
The smaller $\mathbf{H}$ and the consequently simplified logic allows to reduce power consumption to 196.67 mW.

\ctable[
	caption={Proposed decoder implementation results versus \cite{ORBARCH_J, PolarArch}.},
	star,
	label={tab:polar}, 	
    ]{l|rrr|rr|r}{
    \tnote[e]{ORBGRAND algorithm}
    \tnote[f]{$n$=128 only}
    \tnote[g]{SCL algorithm with list size 8}
    \tnote[h]{Single code decoder for PC(128,64) + CRC(6)}
    }{
    	\toprule
              &  \multicolumn{3}{c|}{This work} &  \multicolumn{2}{c|}{\cite{ORBARCH_J}\tmark[e]} & \cite{PolarArch}\tmark[g]\\
	              &  {\bf A$_{\rm PC}$} & {\bf D} & {\bf E} & 1 & 2 &\\
	\midrule
		Result type & \multicolumn{3}{c|}{Synthesis} & \multicolumn{2}{c|}{Synthesis} & Layout \\ 
		Technology [nm]  &  7 & 7 & 7 & 65 & 65 & 28\\
		Supply [V]  &  0.5 & 0.5 & 0.5 & 0.9 & 0.9 & 1.0\\
		$N$ 	 	  &  128 				&	128       &	128     & 128\tmark[f] & 128\tmark[f] &  128\tmark[h]\\
		$R_{min}$   &  1/128 			&	0.75      &	0.75    &  0.75 & 0.75 & 64/128\\
		$r$           &  105/128 			&	105/128      &	105/128    & 105/128 & 105/128  & 64/128\\
		$Q_{max}$	  &  $2^{13}$			&	$2^{13}$  &	$2^{13}$  & $1.77\cdot 2^{16}$ & $1.30\cdot 2^{18}$ & - \\
		$Q_S$		  &  512 				&	512       & 256  & - & - & -\\
		$T$			  &  18				&   18		  & 34 	 & - & - & -\\	
		$Q_{LUT}$	  &  512				&	512   & 512  & - & - & -\\
		$B$         &  8				&  8		&  8	&  5  & 5 & 6\\
			\midrule
 		$f$ [MHz]     &  616				&	616       &  701  & 454 & 454 & 418\\
		Area [mm$^2$] & 5.16 			&   3.38	& 3.70 & 1.82 & 2.25 & 3.15\\
		Area @ 20 nm [mm$^2$] &  		18.58	&  12.168 	& 13.32 & 0.26 & 0.32 & 2.02\\
			\midrule
		B.C. $\mathcal{L}$ ~\multirow{2}{*}{[cc] - [ns]}     & \multirow{2}{*}{25 - 40.58} 	&	\multirow{2}{*}{25 - 40.58} &  \multirow{2}{*}{41 - 58.49}  & 1 - 2.20 & 1 - 2.20 & \multirow{2}{*}{59 - 141.3}\\
		W.C. $\mathcal{L}$    &  	&	 &    &  4223 - 9300 & 93416 - 205760 &\\
			\midrule
		B.C. $\mathcal{T}$ ~\multirow{2}{*}{[$\frac{\rm bits}{\rm cycle}$] - [Gb/s]}   & \multirow{2}{*}{105 - 64.68}  &   \multirow{2}{*}{105 - 64.68}   & \multirow{2}{*}{105 - 73.61} & 105 - 47.67 & 105 - 47.67 & \multirow{2}{*}{64 - 26.75} \\
		W.C. $\mathcal{T}$  & &     & &  0.0248 - 0.0113&  0.0011 - $512{\rm e}^{-6}$ & \\
			\midrule
		Area eff. [Gbps/mm$^2$] & 	12.53     &  19.13  & 19.89 & 23.3 & 18.9 & 8.49\\	
		Area eff. @ 20 nm [$\frac{\rm Gbps}{\rm mm^2}$] & 3.48     & 5.31   &5.53 &183.35 &135.00& 13.24\\	
		Power [mW]  			& 269.10  & 167.26	&170.84  &104.3& 133 & 3340\\
		Energy/bit [pJ/bit] & 4.16	& 2.59 & 2.32  & 2.45& 3.13 & 124.86\\ 

		\hline
}

Implementation {\bf C} instantiates smaller pattern memories, with $Q_S=256$. 
This choice increases the number of decoder stages to $T=34$, resulting in $\mathcal{L}=41$ clock cycles. 
Each stage has lower complexity than the $Q_S=512$ case, thanks to the smaller pattern memory, syndrome calculation circuit, and $\mathbf{z}$ selection networks: the shorter critical path increases the achievable frequency to $f=701$ MHz, and thus the throughput to $\mathcal{T}=79.21$ Gb/s.
On the other hand, the higher $T$ increases the total number of pipeline registers, that do not scale with $Q_S$, leading to an overall area of $4.05$ mm$^2$.
While the smaller $Q_S$ allows to have an even more granular stage activation rate than implementation {\bf A$_{\rm BCH}$} and {\bf B}, the combined effects of the higher frequency and the more numerous pipeline stages bring the power consumption to 204.06 mW.

The flexibility granted by the implementation {\bf A$_{\rm BCH}$} comes at a cost in terms of area efficiency ($\frac{\mathcal{T}}{\rm area}$) and energy per bit ($\frac{\rm power}{\mathcal{T}}$), that are the lowest and highest among the three solutions, respectively. 
Reducing the size of $\mathbf{H}$ allows for 1.39$\times$ improvement in area efficiency and 1.41$\times$ reduction in energy per bit with respect to the fully-flexible implementation.
The higher throughput of implementation {\bf C} allows for an additional 1.04$\times$ factor in area efficiency and 1.10$\times$ energy per bit saving.

The work presented in \cite{GRANDchip} details a fabricated chip that implements the GRAND algorithm \cite{GRAND_first}. 
The generated test patterns are limited to those with $HW\le 3$, and attempted in ascending $HW$ order.
Unlike the proposed architecture, the code length is not fully flexible, and can support only $n=$\{127,128\}.
The decoder in \cite{GRANDchip} has very low power consumption and a small area occupation. 
This is achieved through small internal parallelism, time-sharing of resources, and low $f$; these factors, together with the large $Q_{max}$, also lead to low throughput and a great disparity between the B.C. and W.C. scenarios.
The latency swings between 71 and 29500 clock cycles, and the throughput can be as low as 0.004 bits/cycle. 
These metrics are independent of the technology node, and the proposed architecture yields lower latency and higher throughput than \cite{GRANDchip}.
The low maximum throughput of \cite{GRANDchip} (1.59 bits/cycle or 108 Mb/s) also impacts negatively the efficiency figures: even considering the results scaled to 20 nm technology, the proposed solutions are more efficient than \cite{GRANDchip}.

Table \ref{tab:BCH} also reports synthesis results for the BCH decoder presented in \cite{BCHDEC2}. 
It is a single-code decoder for the BCH(63,51,2) code defined in the wireless body area network standard, implementing a reduced-complexity version of the type-II Chase decoder with 4 test patterns.
While the code length is half of that used in {\bf A$_{\rm BCH}$}, {\bf B}, {\bf C}, and \cite{GRANDchip}, the error-correction capability of the code is the same.
Metrics for a BCH(127,113,2) decoder with the same architecture can be approximated by doubling the area and the latency reported for \cite{BCHDEC2} in Table \ref{tab:BCH}, while frequency remains the same and the throughput is adjusted for the different $r$.
The resulting decoder is smaller than the proposed architecture and \cite{GRANDchip} both, with a latency more than $3\times$ that of {\bf A$_{\rm BCH}$} and a throughput higher than that of \cite{GRANDchip}.

The results reported in Table \ref{tab:polar} consider PC(128,105) + CRC(11) at a target BLER=$10^{-7}$. Implementation {\bf A$_{\rm PC}$} is structurally identical to {\bf A$_{\rm BCH}$}, with updated power, throughput and efficiency due to the different code used, while implementation {\bf D} and {\bf E} have $R_{min}=0.75$ instead of $R_{min}=0.656$ used in {\bf B} and {\bf C}.
These are the same conditions of \cite{ORBARCH_J}, whose results are reported as well in the same table.
The information throughput is lower than before, due to the polar code having a lower $r$ than the BCH(127,113,2). 
The higher SNR (7.4 dB instead of 5.75 dB used in Table \ref{tab:BCH}) allows for generally lower power consumption, since 84\% of received vectors are valid codewords, and leads to an activation rate of stage 2 lower than $0.0001\%$. 
The higher $R_{min}$ considered in implementation {\bf D} and {\bf E} reduces both area occupation and power consumption, and yield improved energy and area efficiency with respect to implementation {\bf B} and {\bf C}.

The work in \cite{ORBARCH_J} presents a decoder architecture for ORBGRAND decoding using a pruned LWO schedule. 
The results labeled under \cite{ORBARCH_J}.1 in Table \ref{tab:polar} limit the $LW$ to 64 and the $HW$ to 6, while those under \cite{ORBARCH_J}.2 consider patterns with $LW\le 96$ and $HW\le 8$, that result in $Q_{max}\approx1.77\cdot 2^{16}$ and $\approx1.30\cdot 2^{18}$, respectively.
Under these conditions, the target BLER$=10^{-7}$ is achieved around SNR$=8$ dB, whereas LA-iLWO yields more than 0.5 dB gain with a substantially smaller $Q_{max}$ at the same BLER.
Much like \cite{GRANDchip}, the architectures in \cite{ORBARCH_J} use time-shared resources to limit complexity and sacrifice W.C. latency and throughput in favor of average measures.
These decoders yield a 1-cycle B.C. latency and a B.C. throughput in bits/cycle as high as that of the proposed decoder, but the high $Q_{max}$ results in very long W.C. latency and extremely low W.C. throughput, orders of magnitude worse than those in  {\bf A$_{\rm PC}$}, {\bf D}, and {\bf E}.
Being the more heavily pruned of the two, and thus having the worst BLER, \cite{ORBARCH_J}.1 yields very good area and energy efficiency.
The \cite{ORBARCH_J}.2 implementation, whose performance matches that of ORBGRAND with the standard LWO schedule, yields instead lower efficiencies.

An unrolled polar decoder architecture for the PC(128,64)+CRC(6) code implementing the SCL decoding algorithm with list size 8 is presented in \cite{PolarArch}, and its implementation results are reported in Table \ref{tab:polar}. 
Similarly to the proposed decoder, the pipelined structure of \cite{PolarArch} yields high throughput (equal to $N\cdot r$ bits/cycle) and low fixed latency, at the cost of large area occupation.

\section{Conclusion}\label{sec:conc}
In this work, the LUT-aided error-pattern schedule for ORBGRAND has been proposed. 
It schedules a number of empirically-observed, highly-likely error patterns before standard error patter generation.
Coupled with the high-performance iLWO schedule, it achieves near-ML performance and the best error-correction performance in literature among ORBGRAND schedules, outperforming more complex GRAND-based decoding algorithms with a smaller number of maximum codebook queries.
The proposed schedule can be easily implemented within decoder architectures in literature.

Making full use of the advantages of the LUT-aided iLWO schedule, a code-agnostic flexible decoder architecture has been proposed. 
It can decode any binary linear block code with code length and rate that fit within constraints set at design time.
Unlike existing GRAND-based decoders, that focus on average performance sacrificing worst case performance, the proposed decoder guarantees fixed high throughput and low latency, making it an attractive choice for latency-constrained applications.
The paradigm shift allows to use the average performance to reduce power consumption.

The proposed architecture has been implemented in 7 nm FinFET technology for various combinations of code and decoder parameters: its fixed throughput and latency outperform the worst-case performance of decoders in literature by orders of magnitude, and also outperform many best-case performance figures.
Depending on implementation parameters, it can achieve a throughput of up to $79.21$ Gb/s with $58.49$ ns latency, and up to $69.61$ Gb/s throughput with $40.58$ ns latency, obtained during the decoding of a BCH code of length 127 and of a polar code of length 128, respectively. 
Given a maximum number number of error patterns, the decoder architecture can be tuned to distribute them among different decoding stages, striking different tradeoffs between power, area occupation, latency, and throughput. 


\end{document}